\documentclass{emulateapj}
\usepackage{multirow}
\usepackage{subfigure}
\def\msun{$M_{\odot}$}

\def\xmm{\textit{XMM-Newton}}
\tabletypesize{\tiny}
\shortauthors{Lin et al.}
\begin{document}

\title{An Ultrasoft X-ray Flare from 3XMM J152130.7+074916: a Tidal Disruption Event Candidate}
\author{Dacheng Lin\altaffilmark{1}, Peter W. Maksym\altaffilmark{2}, Jimmy A. Irwin\altaffilmark{2}, S. Komossa\altaffilmark{3}, Natalie A. Webb\altaffilmark{4,5}, Olivier Godet\altaffilmark{4,5}, Didier Barret\altaffilmark{4,5}, Dirk Grupe\altaffilmark{6}, Stephen D. J. Gwyn\altaffilmark{7}}
\altaffiltext{1}{Space Science Center, University of New Hampshire, Durham, NH 03824, USA, email: dacheng.lin@unh.edu}
\altaffiltext{2}{Department of Physics and Astronomy, University of Alabama, Box 870324, Tuscaloosa, AL 35487, USA}
\altaffiltext{3}{Max-Planck-Institut f\"ur Radioastronomie, Auf dem H\"ugel 69, 53121 Bonn, Germany}
\altaffiltext{4}{CNRS, IRAP, 9 avenue du Colonel Roche, BP 44346, F-31028 Toulouse Cedex 4, France}
\altaffiltext{5}{Universit\'{e} de Toulouse, UPS-OMP, IRAP, Toulouse, France}
\altaffiltext{6}{Space Science Center, Morehead State University, 235 Martindale Drive, Morehead, KY 40351, USA}
\altaffiltext{7}{Canadian Astronomy Data Centre, Herzberg Institute of Astrophysics, 5071 West Saanich Road, Victoria, British Columbia, V9E 2E7, Canada}

\begin{abstract}
We report on the discovery of an ultrasoft X-ray transient source, 3XMM J152130.7+074916. It was serendipitously detected in an \textit{XMM-Newton} observation on 2000 August 23, and its location is consistent with the center of the galaxy \object{SDSS~J152130.72+074916.5} ($z=0.17901$ and $d_L=866$ Mpc). The high-quality X-ray spectrum can be fitted with a thermal disk with an apparent inner disk temperature of 0.17 keV and a rest-frame 0.24--11.8 keV unabsorbed luminosity of $\sim5\times10^{43}$ erg s$^{-1}$, subject to a fast-moving warm absorber. Short-term variability was also clearly observed, with the spectrum being softer at lower flux. The source was covered but not detected in a \textit{Chandra} observation on 2000 April 3, a \textit{Swift} observation on 2005 September 10, and a second \textit{XMM-Newton} observation on 2014 January 19, implying a large variability ($>$260) of the X-ray flux. The optical spectrum of the candidate host galaxy, taken $\sim$11 yrs after the \textit{XMM-Newton} detection, shows no sign of nuclear activity. This, combined with its transient and ultrasoft properties, leads us to explain the source as tidal disruption of a star by the supermassive black hole in the galactic center. We attribute the fast-moving warm absorber detected in the first \textit{XMM-Newton} observation to the super-Eddington outflow associated with the event and the short-term variability to a disk instability that caused fast change of the inner disk radius at a constant mass accretion rate.

\end{abstract}

\section{INTRODUCTION}
\label{sec:intro}
When stars wander too close to the supermassive black holes (SMBHs),
which are believed to be present in the center of most massive
galaxies, they could be disrupted and subsequently accreted, leading
to energetic flares \citep{lioz1979,re1988,re1990}. Such tidal disruption
events (TDEs) provide a unique way to find and study otherwise dormant
SMBHs in galactic nuclei. TDEs involving solar-type stars are expected
to rise very fast (less than a few months) and decay for months to
years, approximately as $t^{-5/3}$, after a period of peak accretion
at a super-Eddington rate. The spectrum is expected to be ultrasoft,
with characteristic temperatures of $\lesssim0.1$ keV, thus mostly in
UV to soft X-rays. The peak X-ray luminosity $L_\mathrm{X}$ can reach around $10^{44}$ erg~s$^{-1}$.  Most candidates discovered thus far belong to such
soft TDEs
\citep[e.g.,][]{grthle1999,kogr1999,koba1999,essako2008,licagr2011,mauler2010,mauler2013,maliir2014,doceco2014}. A
couple of hard TDEs with strong hard X-ray emission (peak isotropic
X-ray luminosity $>$$10^{47}$ erg~s$^{-1}$) were also found
recently and attributed to the presence of a relativistic jet in the
events \citep{blgime2011,bukegh2011,cekrho2012}. We refer to
\citet{ko2012} and \citet{ko2015} for recent reviews of TDEs in X-rays.

There are other X-ray transients reaching X-ray luminosities
comparable to those of TDEs around SMBHs. However, their duration and
lightcurve evolution is typically very different. The X-ray afterglow
of $\gamma$-ray bursts (GRBs) can have peak $L_\mathrm{X}\sim10^{51}$
erg~s$^{-1}$ and last for months \citep[e.g.,][]{letast2014} or, in a
few cases, for years \citep[e.g.,][]{grbuwu2010}. Their X-ray spectra
are generally hard with photon index $\lesssim$2
\citep{grnove2013}. Recently, several ultralong GRBs were discovered,
with the late-time X-ray spectra found to be very soft \citep[photon
  index $>$3, e.g., ][]{pitrge2014,magula2015}. While supernovae (SNe)
seldom have $L_\mathrm{X}$ above $10^{42}$ erg~s$^{-1}$
\citep[e.g.,][]{im2007,lereme2013}, the superluminous supernava
\object{SCP 06F6} was detected in X-rays $\sim$150 days after its
initial discovery, with $L_\mathrm{X}\sim10^{45}$
erg~s$^{-1}$. \citet{sobepa2008} also serendipitously discovered an
extremely luminous X-ray outburst from SN 2008D, with peak
$L_\mathrm{X}\sim6\times10^{43}$ erg~s$^{-1}$. The main event,
however, last only several hundred seconds and could arise from the
prompt shock breakout from the SN progenitor's surface. The outburst
showed a significant hard-to-soft spectral evolution, with photon
index $\sim$1.7, typical of SNe, at the peak, to $\sim$3.2 about 400 s
later.

In our continuing effort to classify X-ray sources serendipitously
detected by \textit{XMM-Newton} and \textit{Chandra}
\citep[e.g.,][]{liweba2012,liweba2014}, we discovered a supersoft X-ray
transient \object{3XMM J152130.7+074916} (J1521+0749 hereafter) in the
\textit{XMM-Newton} Serendipitous Source Catalog \citep[the 3XMM-DR5
  version,][]{rowewa2015}. The source seems to show an X-ray outburst
in 2000. In this paper, we report the properties of this source and
argue that it is probably a soft TDE at a redshift of $z=0.17901$ (the
source luminosity distance $d_L=866$ Mpc, assuming a flat universe
with $H_0$=70 km~s$^{-1}$~Mpc$^{-1}$ and $\Omega_\mathrm{M}$=0.3). In
Section~\ref{sec:reduction}, we describe the data analysis of
\textit{XMM-Newton}, \textit{Chandra}, \textit{Swift} and
\textit{ROSAT} observations. In Section~\ref{sec:res}, we first
identify the host galaxy of our source, followed by the presentation
of its detailed X-ray spectral and timing properties. We discuss the
nature of our source and give conclusions of our study in
Section~\ref{sec:discussion}.

\tabletypesize{\scriptsize}
\setlength{\tabcolsep}{0.02in}
\begin{deluxetable*}{rcccccccc}
\tablecaption{The X-ray Observation Log\label{tbl:obslog}}
\tablewidth{0pt}
\tablehead{\colhead{Obs. ID} &\colhead{Date} & \colhead{Detector} &\colhead{OAA} &\colhead{$T$} &\colhead{$r_\mathrm{src}$}  & \colhead{Count rate} & \colhead{$L_\mathrm{abs}$} & \colhead{$L_\mathrm{unabs}$}\\
 & & & & (ks)& & ($10^{-3}$ counts s$^{-1}$) & \multicolumn{2}{c}{($10^{43}$ erg s$^{-1}$)}\\
(1) & (2) &(3) & (4) & (5) & (6) & (7) & (8) & (9)
}
\startdata
\multicolumn{4}{l}{\xmm:}\\
\hline
0109930101(X1) &2000-08-23 & pn/MOS1/MOS2 & 9.2$\arcmin$ & 31/41/41 & 25$\arcsec$/25$\arcsec$/25$\arcsec$ &$80.6\pm2.7$/$16.1\pm1.1$/$19.5\pm1.2$ & $3.45_{-0.29}^{+0.16}$ & $5.28_{-1.00}^{+2.28}$\\
0723801501(X2) &2014-01-19 & pn/MOS1/MOS2 & 8.6$\arcmin$ & 96/117/117 & 10$\arcsec$/10$\arcsec$/10$\arcsec$ & $<0.17$/$<0.18$/$<0.12$ & $<0.01$ &$<0.02$\\
\hline
\multicolumn{4}{l}{\textit{Chandra}:}\\
\hline
900(C1) & 2000-04-03 & ACIS-I2 &  7.6$\arcmin$ & 57 & 5.1$\arcsec$ & $<0.06$ & $<0.03$ & $<0.04$\\
\hline
\multicolumn{4}{l}{\textit{Swift}:}\\
\hline
00035189001(S1) & 2005-09-10 & XRT & 6.2$\arcmin$ & 9.6 & 20$\arcsec$ & $<0.36$ & $<0.27$ & $<0.42$\\
\hline
\multicolumn{4}{l}{\textit{ROSAT}:}\\
\hline
rp800128n00(R1) & 1992-08-15 & PSPCB & 6.2$\arcmin$ & 9.6 & 30$\arcsec$ & $<0.99$ & $<0.12$ & $<0.19$\\
rh800425n00(R2) & 1994-08-23 & HRI & 6.2$\arcmin$ & 13.4 & 10$\arcsec$  & $<0.41$ & $<0.22$ & $<0.33$ 
\enddata 
\tablecomments{Columns: (1) the observation ID with our designation given in parentheses, (2) the observation start date, (3) the detector, (4) the off-axis angle, (5) the exposures of data used in final analysis, (6) the radius of the source extraction region, (7) the net count rate (0.2--2 keV for \xmm\ observations, 0.3--2 keV for \textit{Swift} and \textit{Chandra} observations, and 0.1--2.4 keV for \textit{ROSAT} observations, all in the observer frame), (8) rest-frame 0.24--11.8 keV luminosity, corrected for Galactic absorption but not intrinsic absorption, (9) rest-frame 0.24--11.8 keV luminosity, corrected for all neutral and ionized absorption. The spectral shape obtained for X1 using the MCD model subject to a fast-moving warm absorber was assumed in calculation of the luminosities for other observations. All uncertainties and upper bounds are at the 90\% confidence level, calculated using the Bayesian approach with the CIAO task \textit{aprates} for all observations except X1.}
\end{deluxetable*}

\tabletypesize{\scriptsize}
\setlength{\tabcolsep}{0.03in}
\begin{deluxetable}{lcccccccccccc}
%\addtolength{\tabcolsep}{-5pt}
\tablecaption{The \textit{XMM-Newton}/OM photometry of J1521+0749 \label{tbl:xmmom}}
\tablewidth{0pt}
\tablehead{
 \colhead{Obs} & \colhead\textit{UVW2} & \colhead\textit{UVM2}  & \colhead\textit{UVW1} &  \colhead\textit{U} &  \colhead\textit{B} &  \colhead\textit{V}
}
\startdata
X1 & \nodata & \nodata & $>20.0$ & $>20.4$ &  $21.5\pm0.5$ & $20.3\pm0.5$\\
X2 & $>18.5$ & $>19.4$ & $>20.4$ & $>20.6$ & $21.2\pm0.4$ & $20.2\pm0.3$
\enddata 
\tablecomments{The AB magnitudes with 1$\sigma$ uncertainties or the 3$\sigma$ detection limits are given. }
\end{deluxetable}

\section{DATA ANALYSIS}
\label{sec:reduction}
Table~\ref{tbl:obslog} lists all the pointed X-ray observations that
covered J1521+0749 and were analyzed by us. There are two
\textit{XMM-Newton} observations, one \textit{Chandra} observation,
one \textit{Swift} observation, and two \textit{ROSAT}
observations. They are denoted as X1, X2, C1, S1, R1 and R2,
respectively, hereafter (refer to the table). All these observations
have the cluster of galaxies MKW 3s as the target, with J1521+0749
observed at off-axis angles of
$\sim$6$\arcmin$--9$\arcmin$. J1521+0749 was serendipitously detected
in X1 but not in the other observations. We extracted the spectrum and
light curve from X1 for detailed study. We also extracted the spectra
and the corresponding response files for the other observations in
order to constrain the luminosity level of our source in these
observations. Circular regions (see Table~\ref{tbl:obslog} for the
radii used) were used to extract the source spectra. Our source is
relatively far away from MKW 3s, but there is still some small
contamination of the X-ray emission from this cluster of galaxies on
our source. In order to accurately represent the contamination level,
we extracted the background spectra from 4--8 circular regions that
are near our source and had the same size and the same distance to MKW
3s as the source region. Below we give more details about the
procedures that we adopted in analysis of each observation.

The source was in the field of view (FOV) of all the three European
Photon Imaging Cameras \citep[i.e., pn, MOS1, and
  MOS2,][]{jalual2001,stbrde2001,tuabar2001} in the imaging mode in
both \textit{XMM-Newton} observations. We used SAS 13.5.0 and the
calibration files of 2014 August for reprocessing the X-ray event
files and follow-up analysis. The data in strong background flare
intervals, seen in all cameras in X1 and in the pn camera in X2, were
excluded following the SAS thread for the filtering against high
backgrounds, i.e., excluding all times when the background level
exceeded the low and steady
level\footnote{http://xmm.esac.esa.int/sas/current/documentation/threads
  /EPIC\_filterbackground.shtml}. The final exposures used are given
in Table~\ref{tbl:obslog}. In the end we used 73\%, 81\%, and 81\% of
the total exposures of pn, MOS1, and MOS2 for X1, respectively. The
corresponding fractions are 97\%, 100\%, and 100\% for X2,
respectively. The event selection criteria that we adopted followed
the default values in the pipeline \citep[see Table~5
  in][]{wascfy2009}. J1521+0749 was also covered in the Optical
Monitor \citep[OM,][]{mabrmu2001} in both observations. In X1, the
\textit{UVW1} (2910 \AA), \textit{U} (3440 \AA), \textit{B} (4500
\AA), and \textit{V} (5430 \AA) filters were used, while in X2, two
more filters, \textit{UVW2} (2120 \AA) and \textit{UVM2} (2310 \AA),
were used. We used the SAS task \textit{omichain} to obtain the
photometry of our source (Table~\ref{tbl:xmmom}).

The C1 observation used the imaging array of the AXAF CCD Imaging
Spectrometer \citep[ACIS; ][]{bapiba1998}. Our source is in the
front-illuminated chip I2. We reprocessed the data to apply the latest
calibration (CALDB 4.5.9) using the script \textit{chandra\_repro} in the
\textit{Chandra} Interactive Analysis of Observations (CIAO, version 4.6)
package. We used a source extraction region enclosing
70\% of the point spread function.

In the S1 observation, the X-ray telescope \citep[XRT;][]{buhino2005}
was operated in Photon Counting mode for 9.6 ks, and the UV-Optical
Telescope \citep[UVOT;][]{rokema2005} used the \textit{UVW1} filter
(2910 \AA) for 8.9 ks. We analyzed the data with FTOOLS 6.16 and the
latest calibration files (XRT 20140730 and UVOT 20130118). The X-ray
data were reprocessed with the task \textit{xrtpipeline} (version
0.13.1) to update the calibration. Other than estimating the
luminosity level of our source in the XRT, we also calculated the UV
emission level with the task \textit{uvotsource}, using radii of
5$\arcsec$ and 20$\arcsec$ for the circular source and background
regions, respectively.

The \textit{ROSAT} observations were also analyzed with FTOOLS
6.16. We note that there are three more \textit{ROSAT} pointed
observations that covered J1521+0749. We did not include them in this
study because they are not deep enough to provide good constraints on
the flux limits of our source, due to large off-axis angles and/or
short exposures. J1521+0749 was not detected either in these
observations.

\begin{deluxetable}{lccc}
\tablecaption{Fitting results of the X1 spectrum from J1521+0749\label{tbl:spfit}}
\tablewidth{0pt}
\startdata
\hline
Models & MCD+Ga\tablenotemark{a} & BB+Ga\tablenotemark{a}\\
$N_\mathrm{H,i}$ (10$^{20}$ cm$^{-2}$) & $0.3^{+1.4}$ & $  0.0^{+0.4}$\\
$kT_\mathrm{MCD}$/$kT_\mathrm{BB}$ (keV)& $0.125^{+0.005}_{-0.006}$  & $0.099^{+0.002}_{-0.003}$\\
$N_\mathrm{MCD}$/$N_\mathrm{BB}$ &$  216^{+  101}_{  -46}$ & $  629^{+  106}_{  -76}$\\
$E_\mathrm{ga}$ (keV)& $ 0.75^{+ 0.02}_{-0.07}$ & $ 0.76^{+ 0.02}_{-0.03}$ \\
$\sigma_\mathrm{ga}$ (keV)& $0.02^{+0.06}$ & $0.00^{+0.05}$\\
$N_\mathrm{ga}$ ($10^{-5}$)& $ 1.0^{+ 1.1}_{-0.5}$ & $ 0.8^{+ 0.5}_{-0.3}$\\
$L_\mathrm{abs}$ (10$^{43}$ erg~s$^{-1}$)\tablenotemark{b}  & $ 3.48^{+ 0.20}_{-0.29}$  & $ 3.20^{+ 0.13}_{-0.13}$\\
$L_\mathrm{unabs}$ (10$^{43}$ erg~s$^{-1}$)\tablenotemark{c} & $ 3.65^{+ 0.58}_{-0.24}$  & $ 3.20^{+ 0.17}_{-0.13}$\\
$L_\mathrm{bol}$ (10$^{43}$ erg~s$^{-1}$)\tablenotemark{d} & $ 7.40^{+ 1.64}_{-0.63}$  & $ 4.24^{+ 0.26}_{-0.21}$\\
$C/\nu$\tablenotemark{e} & 536.4/528 & 545.9/528 \\
\hline
Models & edge*MCD &edge*BB\\
$N_\mathrm{H,i}$ (10$^{20}$ cm$^{-2}$) & $0.0^{+0.6}$ & $0.0^{+0.4}$\\
$kT_\mathrm{MCD}$/$kT_\mathrm{BB}$ (keV) & $0.144^{+0.008}_{-0.006}$  & $0.106^{+0.004}_{-0.003}$\\
$N_\mathrm{MCD}$/$N_\mathrm{BB}$ &$  105^{+   28}_{  -20}$ & $  458^{+   81}_{  -74}$\\
$E_\mathrm{edge}$ (keV) &$ 0.85^{+ 0.02}_{-0.02}$  &$ 0.84^{+ 0.03}_{-0.03}$\\
$\tau_\mathrm{edge}$ & $ 0.98^{+ 0.35}_{-0.32}$ & $ 0.52^{+ 0.32}_{-0.30}$\\
$L_\mathrm{abs}$ (10$^{43}$ erg~s$^{-1}$)\tablenotemark{b} & $ 3.35^{+ 0.15}_{-0.15}$  & $ 3.07^{+ 0.13}_{-0.13}$\\
$L_\mathrm{unabs}$ (10$^{43}$ erg~s$^{-1}$)\tablenotemark{c} & $ 3.46^{+ 0.21}_{-0.13}$  & $ 3.13^{+ 0.13}_{-0.12}$\\
$L_\mathrm{bol}$ (10$^{43}$ erg~s$^{-1}$)\tablenotemark{d} & $ 6.32^{+ 0.55}_{-0.36}$  & $ 4.04^{+ 0.21}_{-0.20}$\\
$C/\nu$\tablenotemark{e} &529.7/529 & 552.8/529 \\
\hline
Models & zxipcf*MCD &zxipcf*BB\\
$N_\mathrm{H,i}$ (10$^{20}$ cm$^{-2}$) & $  0.0^{+  1.6}$ & $  0.0^{+  0.6}$\\
$kT_\mathrm{MCD}$/$kT_\mathrm{BB}$ (keV) & $0.169^{+0.024}_{-0.017}$  & $0.130^{+0.007}_{-0.007}$\\
$N_\mathrm{MCD}$/$N_\mathrm{BB}$ &$   72^{+   36}_{  -20}$ & $  577^{+   97}_{  -78}$\\
$N_\mathrm{H,zxipcf}$ (10$^{22}$ cm$^{-2}$) & $  4.7^{+  2.7}_{ -1.2}$ & $  2.8^{+  1.3}_{ -0.6}$\\
$\log\xi_\mathrm{zxipcf}$& $ 2.02^{+ 0.10}_{-0.09}$ & $ 1.29^{+ 0.09}_{-0.09}$\\
$z_\mathrm{zxipcf}$& $-0.12^{+ 0.02}_{-0.02}$ & $-0.12^{+ 0.02}_{-0.03}$\\
$L_\mathrm{abs}$ (10$^{43}$ erg~s$^{-1}$)\tablenotemark{b} & $ 3.45^{+ 0.16}_{-0.29}$  & $ 3.20^{+ 0.14}_{-0.14}$\\
$L_\mathrm{unabs}$ (10$^{43}$ erg~s$^{-1}$)\tablenotemark{c} & $ 5.28^{+ 2.28}_{-1.00}$  & $ 9.94^{+ 3.67}_{-2.56}$\\
$L_\mathrm{bol}$ (10$^{43}$ erg~s$^{-1}$)\tablenotemark{d} & $ 8.6^{+ 3.2}_{-1.3}$  & $11.6^{+ 4.1}_{-2.8}$ \\
$C/\nu$\tablenotemark{e} &523.8/528 & 532.1/528 \\
\hline
Models & zxipcf*optxagnf$_\mathrm{Sch}$\tablenotemark{f} &zxipcf*optxagnf$_\mathrm{Kerr}$\tablenotemark{g}\\
$N_\mathrm{H,i}$ (10$^{20}$ cm$^{-2}$) & $  0.0^{+  1.6}$ & $  0.0^{+  1.6}$\\
$M_\mathrm{BH}$ (\msun) & $ 1.9^{+ 0.4}_{-0.4}\times10^{5}$& $1.4^{+ 0.3}_{-0.3}\times10^{6}$ \\
$L_\mathrm{Bol}/L_\mathrm{Edd}$ & $ 4.2^{+ 1.9}_{-1.0}$ & $ 0.51^{+ 0.28}_{-0.12}$\\
$N_\mathrm{H,zxipcf}$ (10$^{22}$ cm$^{-2}$) & $  4.7^{+  2.7}_{ -2.3}$ & $  4.7^{+  2.7}_{ -2.3}$\\
$\log\xi_\mathrm{zxipcf}$& $ 2.02^{+ 0.15}_{-0.09}$ & $ 2.02^{+ 0.13}_{-0.07}$\\
$z_\mathrm{zxipcf}$& $-0.12^{+ 0.02}_{-0.02}$ & $-0.12^{+ 0.02}_{-0.03}$\\
$C/\nu$\tablenotemark{e} & 523.7/528 & 523.7/528
\enddata 
\tablecomments{All fits used the C statistic. \textsuperscript{a}``Ga'' is a Gaussian emission line; \textsuperscript{b}rest-frame 0.24--11.8 keV luminosity, corrected for Galactic absorption but not intrinsic absorption; \textsuperscript{c}rest-frame 0.24--11.8 keV luminosity, corrected for all neutral and ionized absorption; \textsuperscript{d}the bolometric luminosity based on the total flux of the thermal component (MCD or BB); \textsuperscript{e}the C statistic and the degrees of freedom; \textsuperscript{f}for thermal emission from an accretion disk around a non-rotating Schwarzschild BH;  \textsuperscript{g}for thermal emission from an accretion disk around a maximally rotating Kerr BH.}
\end{deluxetable}

\section{RESULTS}
\label{sec:res}
\subsection{The X-ray Source Position and the Host Galaxy}
\label{sec:multiwav}
\begin{figure}[!tb]
\centering
\includegraphics[width=3.4in]{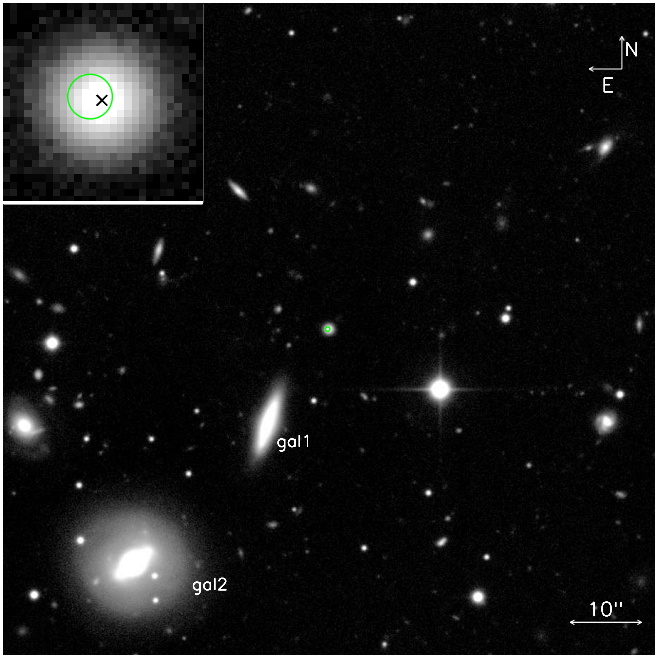}
\caption{The CFHT/MegaPrime $r\arcmin$-band image around the field of
  J1521+0749. The green circle of radius 0\farcs58 (i.e., 1.8
  kpc) represents the 99\% positional uncertainty of J1521+0749,
  indicating that J1521+0749 is consistent with the center of the
  galaxy SDSS J152130.72+074916.5. The two closest galaxies to
  J1521+0749 in MKW 3s are SDSS J152131.80+074851.9 and SDSS
  J152134.15+074815.3 ('gal1' and 'gal2' in the plot,
  respectively). The inset plot in the upper left corner zooms in on
  the $5\arcsec\times5\arcsec$ region around SDSS
  J152130.72+074916.5. The center of the galaxy is marked with a
  cross.  \label{fig:cfhtimg}}
\end{figure}

The position of J1521+0749 in the 3XMM-DR5 catalog is only $0\farcs06$
away from the Sloan Digital Sky Survey \citep[SDSS,][]{abadag2009}
galaxy \object{SDSS~J152130.72+074916.5}
(Figure~\ref{fig:cfhtimg}). However, such a small offset could be an
artifact to some extent, considering that this source was among those
used for the absolute astrometry correction and that it is one of the
brightest sources in X1. Therefore, we re-calculated the source
position. We first obtained the mean positions (weighted by the
uncertainties) of the point sources in X1, X2 and C1, after correcting
the astrometry of X2 and C1 relative to X1 using the SAS task
\textit{catcorr}. For C1, we included a systematic uncertainty of
0\farcs16 obtained in \citet{robu2011}. The task \textit{catcorr} also
calculated the systematic positional uncertainties from the
astrometric rectification procedure \citep{rowewa2015}. Such
systematic uncertainties were taken into account and propagated. The
mean positions, excluding J1521+0749 and those in bright X-ray
emission from MKW 3s, were then cross-correlated with optical sources
to obtain the absolute astrometry correction. For the optical sources,
we stacked the Canada-France-Hawaii Telescope (CFHT) MegaPrime/MegaCam
\citep{bochab2003} images (aligned to the SDSS astrometry) in the
$r\arcmin$ band taken between 2008 April and 2010 April using MegaPipe
\citep{gw2008}, resulting in a total exposure of 3840 s and a median
seeing FWHM of 0\farcs77. The new absolute astrometry correction gave
the position of J1521+0749 to be R.A.=15:21:30.75 and
Decl.=+7:49:16.7, with the 99\% positional uncertainty of 0\farcs58
(i.e., 1.8 kpc at the redshift $z=0.17901$ of the candidate
  host galaxy of J1521+0749; see below). The relatively small
positional uncertainty is mostly thanks to the very high quality of
the X1 detection. This new position is 0\farcs33 away from the center
of SDSS J152130.72+074916.5, consistent within the uncertainty
(Figure~\ref{fig:cfhtimg}). The number density of the optical sources
that are as bright as or brighter than SDSS J152130.72+074916.5 in the
$r\arcmin$ band within $10\arcmin$ is 0.00038 per square
arcsec. Thus the chance probability for our X-ray source to be
  within $0\farcs33$ from the center of SDSS J152130.72+074916.5 is
  only 0.0001.

\begin{figure}
\hspace*{-0.2in}\includegraphics[width=3.7in]{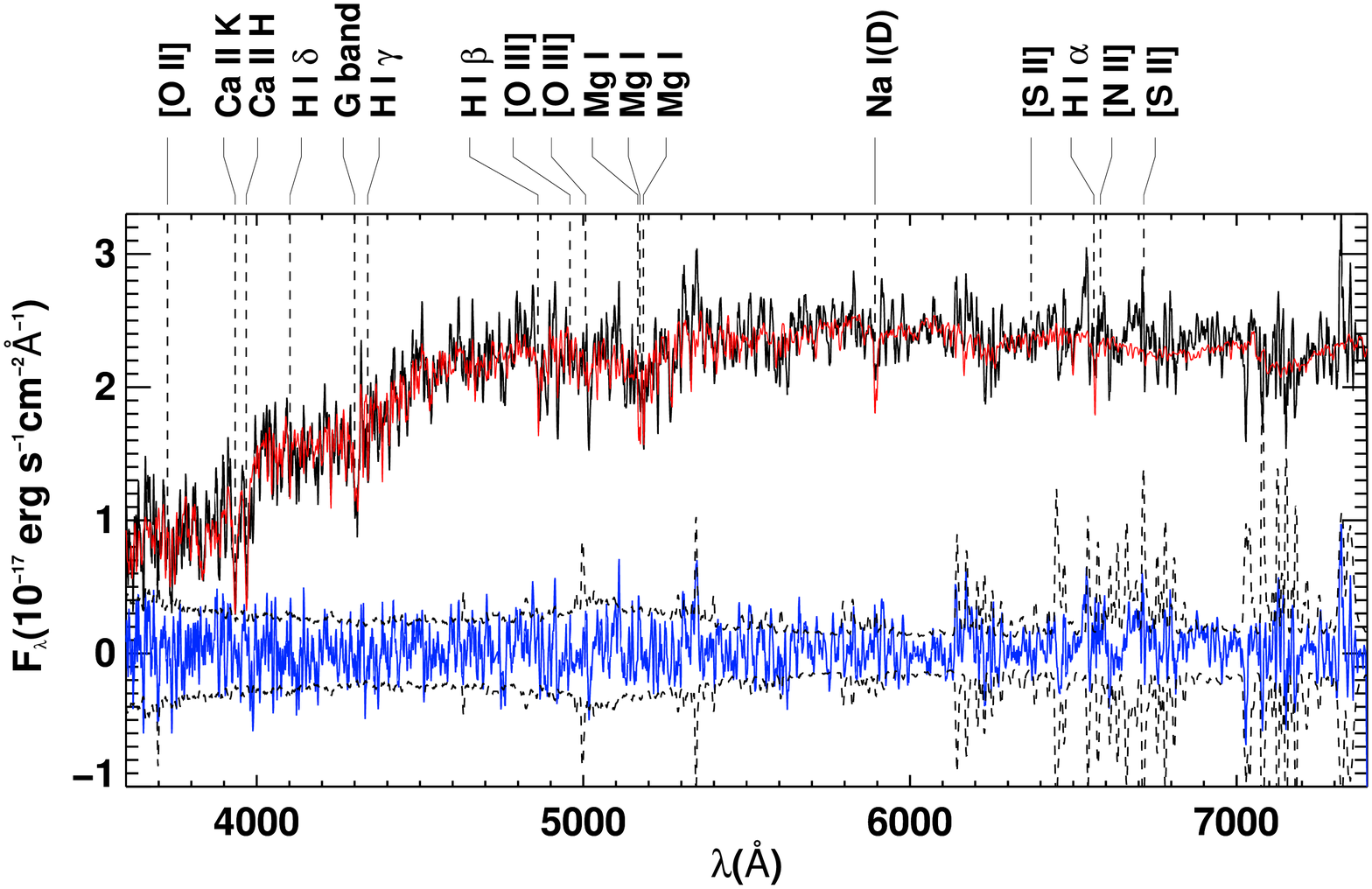}\vspace{-0.1in} \\
\hspace*{-0.2in}\includegraphics[width=3.7in]{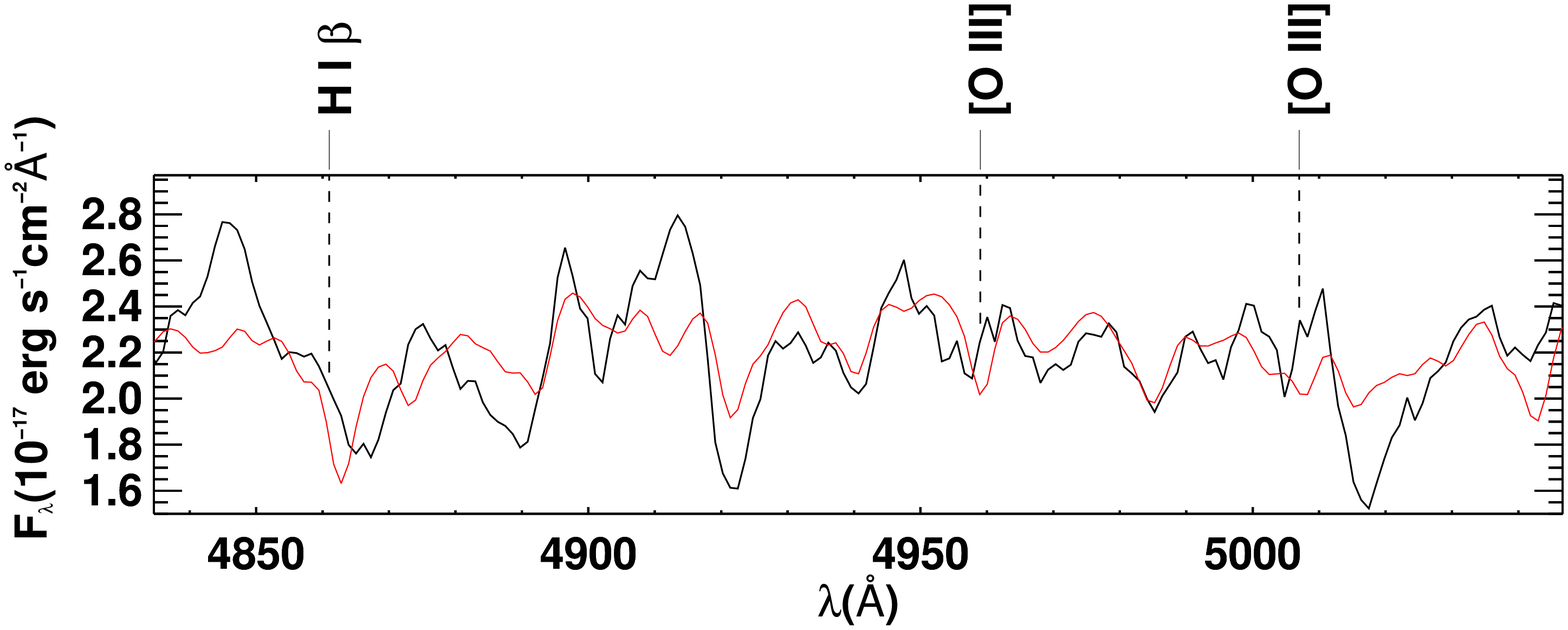} \\
%\hspace*{0.13in}
\hspace*{-0.05in}\includegraphics[width=3.5in]{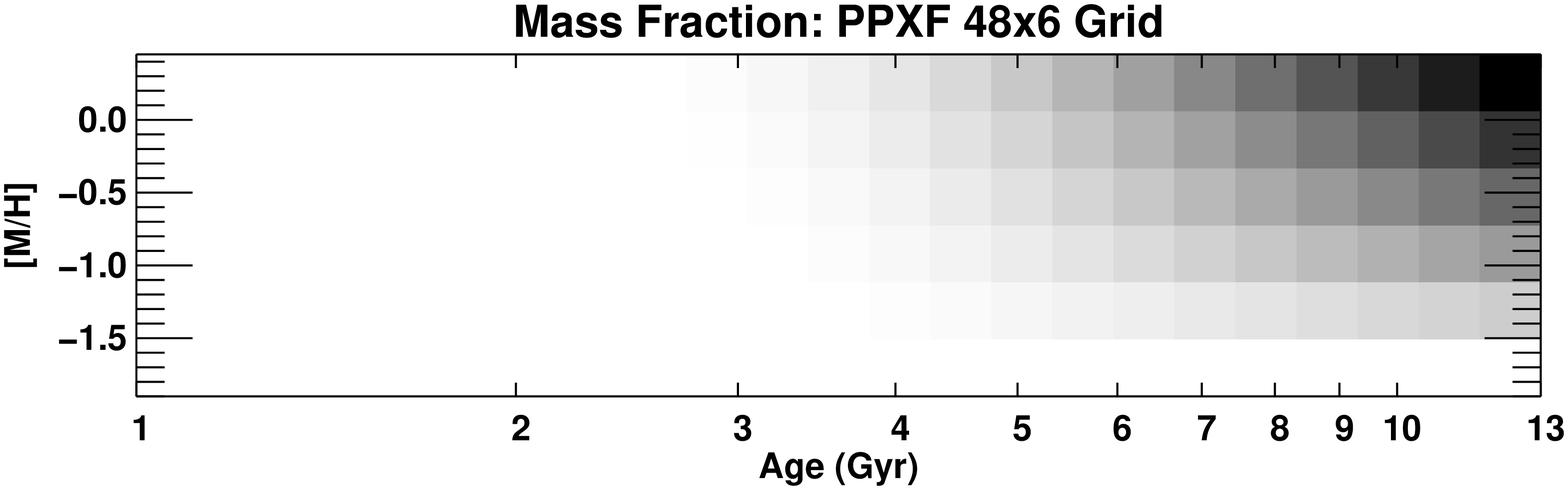} \\
\caption[]{Top panel: PPXF best-fit modeling of J1521+0749.  The SDSS rest-frame data (black, solid) are overplot with the best-fit PPXF model (red).  Points are smoothed with a boxcar function over 5 pixels for clarity.  Smoothed residuals are also plotted (blue) with mean noise (black, dashed).  A strong \ion{O}{1} $\lambda5577\;$\AA\ (observed) sky line has been masked out.  Important stellar absorption and AGN diagnostic emission lines are overplot for reference.  Note that several strong false `features" are associated with atmospheric OH complexes at long $\lambda$, particularly near H$\alpha$ $\lambda6563\;$\AA. Middle panel: As with the top, with the region about [\ion{O}{3}] expanded for clarity and emphasis. Bottom panel: Relative mass fractions of different stellar populations with respect to metallicity and age, as per the best-fit model the {\small PPXF} template synthesis code.  Dominant populations are indicated by shading, with darker shading indicating a larger fraction of the total stellar mass in the best-fit model.}
\label{fig:s13s2740gsp}
\end{figure}

The SDSS took a spectrum of the galaxy on 2011 May 28, which is shown
in Figure~\ref{fig:s13s2740gsp} (from the SDSS archive). Although the
spectrum is a little noisy, the typical absorption features indicating
a non-active galaxy can be seen. The redshift of the galaxy is
$z=0.17901\pm0.00007$ ($D_L=866$ Mpc). No clear emission lines, except
some spikes due to contamination from sky lines, can be seen. The
$3\sigma$ upper limit of the flux of [\ion{O}{3}] $\lambda$5007 is
$1.8\times10^{-17}$ erg s$^{-1}$ cm$^{-2}$, corresponding to a
luminosity of $1.6\times10^{39}$ erg s$^{-1}$ (Galactic extinction
corrected). Using the bolometric correction factors from the
[\ion{O}{3}] $\lambda$5007 flux in \citet{labima2009}, we obtained the
$3\sigma$ upper limit of the bolometric luminosity of the persistent
nuclear activity to be $1.4\times10^{41}$ erg s$^{-1}$.

In order to examine the host galaxy properties in greater detail, we
fitted the SDSS spectrum to multi-component models comprised of
single-population synthetic spectra, as in \cite{maulro2014}.  Using
Penalized Pixel Fitting ({\tt PPXF}) software \citep{caem2004} and
\cite{vasafa2010} synthetic spectra spanning a grid of 48 ages up to
14 Gyr, with [M/H]=\{$+0.22$, $0.00$, $-0.40$, $-0.71$, $-1.31$,
$-1.71$\}, we found that J1521+0749 is consistent with an old
($\gtrsim10\;$Gyr), passive stellar population. Full-spectrum
kinematics implied a stellar dispersion
$\sigma_\star\sim66\;\rm{km\;s}^{-1}$, which is comparable to the
limit from SDSS instrumental dispersion
$\sigma_\star\sim69\;\rm{km\;s}^{-1}$. We likewise modeled the
spectrum with {\tt STARLIGHT} and two arrays of \cite{brch2003}
synthetic spectra, as in \cite{maulro2014}. {\tt STARLIGHT} inferred a
stellar mass $M_\star\sim8.4\times10^9\;$\msun. Care should be taken
in interpreting $\sigma_\star$, since in general $\rm{S/N}\lesssim10$
for $\lambda\lesssim5400\;$\AA (rest-frame). Using \cite{grsc2015}, we
inferred the central BH mass $M_\mathrm{BH}\sim2\times10^7\;$\msun,
which has a $1\sigma$ uncertainty of 0.83 dex. The BH mass can also be
estimated using the BH mass versus bulge $K$-band luminosity relation
\citep{gr2007,mahu2003} and is $\lesssim5\times10^7$ \msun, based on
the $K$-band magnitude of $m_K=16.99$ ($M_K=-22.4$) of the galaxy from
the the 3.8m United Kingdom Infra-red Telescope (UKIRT) Infrared Deep
Sky Survey \citep{lawaal2007}. These BH mass estimates are consistent
with each other and are a little larger than the estimate from the
X-ray spectral fit in Section~\ref{sec:spmodel} ($\sim$$10^5$--$10^6$
\msun).

\subsection{The Long-term Variability}
\begin{figure}[!tb]
\centering
\includegraphics[width=3.4in]{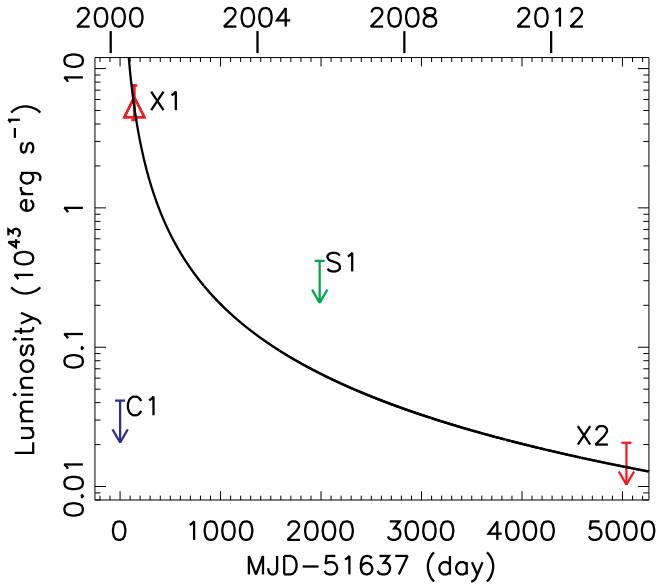}
\caption{The long-term unabsorbed luminosity curve in the rest-frame
  0.24--11.8 energy band. The uncertainties and upper bounds are at the 90\% confidence
  level (see Table~\ref{tbl:obslog}). The observations are noted in
  the plot. The solid line represents a $(t-t_\mathrm{D})^{-5/3}$
  decline, assuming that X1 is on it and that the disruption time
  $t_\mathrm{D}$ is at the time of C1 (i.e., 4.7 months before
  X1). \label{fig:lumlc}}
\end{figure}

\begin{figure}[!tb]
\centering
\includegraphics[width=3.4in]{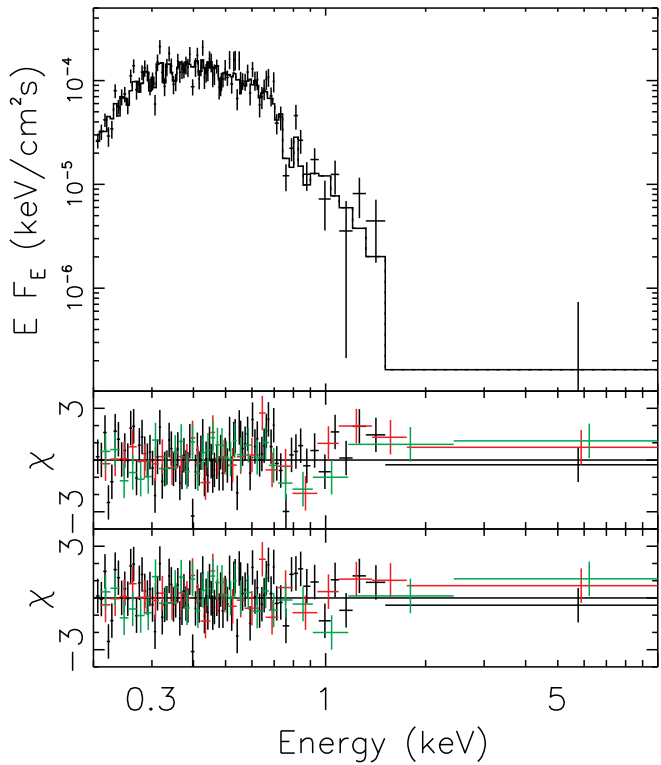}
\caption{(top panel) The unfolded spectrum of X1 (for clarity, only
  the pn spectrum is shown). The spectrum was fitted with the MCD
  model subject to a fast-moving warm absorber (solid line). (middel
  panel and bottom panel) The fit residuals with the MCD model without
  and with the warm absorber, respectively (pn in black, MOS1 in red,
  and MOS2 in green). For clarity, the spectrum was rebinned to be
  above 1$\sigma$ per bin in the plot. \label{fig:srcspec}}
\end{figure}

Figure~\ref{fig:lumlc} plots the long-term evolution of the unabsorbed
luminosity $L_\mathrm{X}$ of J1521+0749 in its rest-frame 0.24--11.8
keV energy band (or the observer-frame 0.2--10 keV energy band). The
luminosities were obtained based on the fit to X1 with a multicolor
disk (MCD) subject to a fast-moving warm absorber
(Section~\ref{sec:spmodel}) and adopting a distance of $D_L=866$ Mpc
(Section~\ref{sec:multiwav}). The source was not detected in
observation C1 in 2000 April, with $L_\mathrm{X}<4.1\times10^{41}$
(the 90\% confidence upper bound; the same below;
Table~\ref{tbl:obslog}). The source was detected in X1 4.7 months
later, with $L_\mathrm{X}\sim5.3\times10^{43}$ erg~s$^{-1}$. In S1 in
2005 September, the source was not detected, with
$L_\mathrm{X}<4.2\times10^{42}$ erg~s$^{-1}$. It was not detected
either in X2 in 2014 January, with $L_\mathrm{X}<2.1\times10^{41}$
erg~s$^{-1}$. Therefore the source has a long-term variability factor
of $\gtrsim$260. Figure~\ref{fig:lumlc} does not include the
\textit{ROSAT} observations. We estimated $L_\mathrm{X}$ in R1 and R2
to be $<1.9\times10^{42}$ erg~s$^{-1}$ and $<3.3\times10^{42}$
erg~s$^{-1}$, thus a factor of $>$28 and $>$16 lower than that in X1,
respectively.

In the UV and optical, the source was detected in the B and V filters,
but not in other filters, in X1 and X2 (Table~\ref{tbl:xmmom}). The
optical fluxes in these two \textit{XMM-Newton} observations are
consistent with each other and the SDSS measurements
($u\arcmin=22.4\pm0.4$ mag, $g\arcmin=21.09\pm0.05$ mag,
$r\arcmin=19.96\pm0.02$ mag, $i\arcmin=19.51\pm0.02$ mag, and
$z\arcmin=19.17\pm0.06$ mag). In the \textit{UVW1} filter in S1, there
is no clear UV emission from J1521+0749 either, except some
contamination from the read-out streak of a bright source. Although we
obtained the \textit{UVW1} AB magnitude of 23.3$\pm$0.3 mag for
J1521+0749 using \textit{uvotsource}, this value should be viewed as
an upper limit. Therefore we observed no clear UV/optical variability
accompanying the X-ray flare detected in X1.

\subsection{Modeling of the X-ray Spectrum in X1}
\label{sec:spmodel}
We fitted the X1 spectrum with several common simple models. To check
the quality of the fits, we rebinned the spectrum to have a minimum of
20 count per bin and adopted the $\chi^2$ statistic in the
fits. However, all the best-fitting parameters that we will present
were obtained from the final fits to the spectrum that were rebinned
to have a minimum of one count per bin and adopted the C statistic,
considering that the spectrum most likely include narrow absorption
lines. The fits using the C statistic gave smaller uncertainties of
the parameters than those using the $\chi^2$ statistic, but they are
consistent with each other within the uncertainties. Because our
source is most likely associated with
\object{SDSS~J152130.72+074916.5} at $z=0.17901$
(Section~\ref{sec:multiwav}), we applied this redshift to the spectral
models with the convolution model \textit{zashift} in XSPEC. All
models included the Galactic absorption which we fixed at
$N_\mathrm{H}=2.61\times10^{20}$ cm$^{-2}$ \citep{kabuha2005} using
the \textit{tbabs} model. We also included possible absorption
intrinsic to the source using the \textit{ztbabs} model. The abundance
tables of \citet{wialmc2000} were used.

We first fitted the spectrum with a power law (PL) and obtained an
unphysically high photon index of $\Gamma_\mathrm{PL}=5.9\pm0.2$ (the
uncertainties from the spectral fits throughout the paper are all at
the 90\% confidence level), indicating an ultrasoft source. However,
the fit is not acceptable, with the reduced $\chi^2$ value of
$\chi^2_\nu=1.63$ for degrees of freedom $\nu=147$. The fits with a
MCD model (\textit{diskbb} in XSPEC) is much better, with
$\chi^2_\nu=1.18$. However, clear residuals are seen between 0.5--1
keV (middle panel in Figure~\ref{fig:srcspec}). We investigated
several scenarios to explain the residuals. We first tried to add a
PL. The fit was only marginally improved (at the 90\% significance
level), if we fixed $\Gamma_\mathrm{PL}$ at 2.0 and 3.0, with the PL
contributing $<2$\% and $<4$\% ($2\sigma$ upper limit) of the
rest-frame 0.24--11.8 keV unabsorbed luminosity, respectively. We next
tried the emission line explanation for the residuals by adding a
Gaussian line (\textit{gaussian} in XSPEC) to the MCD model. We
obtained a significantly improved fit with $\chi^2_\nu=1.09$ for
$\nu=144$ (i.e., a total decrease of $\chi^2$ of 17.8 for three more
degrees of freedom compared with the MCD fit). The final fit using the
C statistic is given in Table~\ref{tbl:spfit}. The line has a centroid
rest-frame energy $E_\mathrm{ga}=0.75_{-0.07}^{+0.02}$ keV and a width
$\sigma_\mathrm{ga}=0.02^{+0.06}$ keV, and the apparent inner disk
temperature is $kT_\mathrm{MCD}=0.125\pm0.006$ keV. The neutral
absorption is negligible, which is also the case for all models tested
below.

We then checked whether the residuals could be due to ionized
absorption by applying the \textit{edge} model on the MCD model and
obtained a good fit with $\chi^2_\nu=1.05$ ($\nu=145$), i.e. $\chi^2$
reduced by 22.5 (or the C statistic reduced by 27.2) for two more
degrees of freedom. The final fit using the C statistic is given in
Table~\ref{tbl:spfit}. The edge energy is
$E_\mathrm{edge}=0.85\pm0.02$ keV (rest-frame), and optical depth is
$\tau_\mathrm{edge}=1.0\pm0.3$. The apparent inner disk temperature is
now $kT_\mathrm{MCD}=0.144\pm0.007$ keV. To obtain the confidence
level of the improvement of the fit from adding the \textit{edge}
model, we applied the posterior predictive $p$-value method
\citep{huvaos2008,prvaco2002}. With 10,000 spectra simulated, we found
none having the reduction of the C statistic from the introduction of
the \textit{edge} model to be as large as obtained for the X1 spectrum
(i.e., 27.2). Therefore the chance probability of improving the fit by
adding the \textit{edge} model is $<10^{-4}$.

We tried to replace the \textit{edge} model with the more physical
warm absorber model \textit{zxipcf} \citep{redopo2008}. We obtained a
good fit with $\chi^2_\nu=1.04$ ($\nu=144$). The final fit using the C
statistic is given in Table~\ref{tbl:spfit}. The warm absorber has
$N_\mathrm{H}=(5_{-1}^{+3})\times10^{22}$ cm$^{-2}$, the ionization
parameter $\log\xi=2.0\pm0.1$, and an outflow line-of-sight velocity
of $(0.12\pm0.02) c$, where $c$ is the speed of light. The apparent
inner disk temperature is slightly higher than inferred from the
models investigated above ($kT_\mathrm{MCD}=0.17\pm0.02$ keV). The fit
and the corresponding residuals, using the $\chi^2$ statistic, are
shown in the top and bottom panels in Figure~\ref{fig:srcspec},
respectively. For this model, we checked the possible presence of a PL
component and found it to improve the fit only at the 90\% confidence
level and contribute $<4$\% and $<12$\% ($2\sigma$ upper limit) of the
rest-frame 0.24--11.8 keV unabsorbed luminosity, if we fixed
$\Gamma_\mathrm{PL}$ at 2.0 and 3.0, respectively.

Ultrasoft X-ray spectra are also often fitted with a
single-temperature blackbody (BB, \textit{bbodyrad} in XSPEC), and our
fits with this model are given in Table~\ref{tbl:spfit}. Because the
residuals in 0.5--1 keV seen in the MCD fit are also present in the BB
fit, we also tried a Gaussian line, an edge and a warm absorber to account
for them. Generally, the fits using a BB have a $\chi^2$ value higher
by 4--8 than those using a MCD. The BB temperatures inferred are in
the range between 0.10--0.13 keV.

Finally, we roughly estimated the BH mass with the AGN spectral model
\textit{optxagnf} (in XSPEC) by \citet{dodaji2012}, assuming that the
X1 spectrum is due to pure thermal disk emission (subject to a warm
absorber). We note that \textit{optxagnf} was created assuming a disk
inclination of $60\degr$. Table~\ref{tbl:spfit} gives the results of
two fits, one for a non-rotating Schwarzschild BH and the other for a
maximally rotating Kerr BH. The former inferred the BH mass of
$(1.9\pm0.4)\times10^{5}$ \msun\ and the Eddington ratio of
$4.2^{+1.9}_{-1.0}$, while the latter inferred the mass of
$(1.4\pm0.3)\times10^{6}$ \msun\ and the Eddington ratio of
$0.51^{+0.28}_{-0.12}$. The quality of both fits is as good as that we
obtained with the MCD model, as indicated by the $\chi^2$ value.

\subsection{The Short-term Variability in X1}
\label{sec:stvar}
\begin{figure}[!tb]
\centering
\includegraphics[width=3.4in]{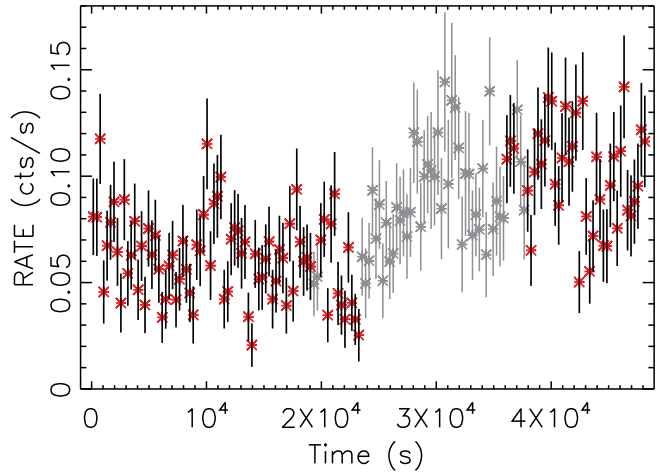}
\caption{The observer-frame 0.2--2 keV pn background subtracted
  light curve from X1 created with the SAS tool \textit{epiclccorr}. The
  bin size is 300~s. Data in the strong flare background intervals in
  the middle of the observation are shown in gray and seem to smoothly
  connect the data before and after the background
  flare.  \label{fig:srclc}}
\end{figure}

Figure~\ref{fig:srclc} shows the observer-frame 0.2--2 keV pn light
curves binned at 300 s from X1. The source is clearly brighter at the
end of the observation than at the beginning. The probability that the
count rate is constant (excluding the high background flaring period)
is $10^{-14}$, from the $\chi^2$ test. There may be fast
variations by a factor of $\sim$2 on timescales of thousands of
seconds, such as those at $2.2\times10^4$~s and $4.3\times10^4$~s into
the observation. We extracted a low-state spectrum from the first 28
ks of the observation and a high-state spectrum from the last 14 ks of
the observation, excluding data in the high background intervals, and
obtained the 0.6--1.2 keV to 0.2--0.6 keV pn count rate ratio to be
$0.157\pm0.015$ and $0.262\pm0.021$, respectively. Therefore the
spectrum is softer at lower flux, at the $4.1\sigma$ significance level.

We fitted these low-state and high-state spectra with the MCD model
subject to a fast-moving warm absorber
(Section~\ref{sec:spmodel}). The chance probability of
  improving the fit by adding the warm absorber is still low, 0.0003 and 0.001 for the low-state and
  high-state spectra, respectively, based on the posterior predictive
  $p$-value method using 10,000 simulations for each spectrum. The parameters of the neutral and ionized
absorption were consistent with and thus fixed at values inferred from
the fit to the average spectrum of the whole observation. We obtained
the disk parameters $kT_\mathrm{MCD}=0.156\pm0.008$ keV and
$N_\mathrm{MCD}=93_{-19}^{+23}$ for the low-state spectrum and
$kT_\mathrm{MCD}=0.189\pm0.010$ keV and
$N_\mathrm{MCD}=60_{-12}^{+16}$ for the high-state spectrum. The disk
temperatures differ by 33 eV at the 4.3$\sigma$ confidence level,
while the inner disk radii differ by 24\% at the 2.2$\sigma$
confidence level. The mass accretion rates, ${\dot M} \propto
T_\mathrm{MCD}^4N_\mathrm{MCD}^{3/2}$ for the MCD model, differ by
11\%, less than 1$\sigma$ uncertainty (we used the XSPEC command
\textit{steppar} on $T_\mathrm{MCD}$ and $N_\mathrm{MCD}$ to calculate
the uncertainty of ${\dot M}$). Therefore one explanation for the
short-term variability in X1 is fast change of the inner disk radius
at an approximately constant mass accretion rate into the
disk. Figure~\ref{fig:srclc} shows that the change occurred within
$\sim$10 ks.

\section{DISCUSSION and CONCLUSIONS}
\label{sec:discussion}
J1521+0749 is an ultrasoft X-ray transient consistent with the center
of a galaxy at $z=0.17901$. This galaxy is consistent with an old
($\gtrsim10\;$Gyr), passive stellar population, based on our modeling
of its SDSS spectrum, which shows no clear emission lines. The
$3\sigma$ upper limit of the bolometric luminosity of the persistent
nuclear activity estimated from the [\ion{O}{3}] upper limit is nearly
three orders of magnitude lower than the peak X-ray luminosity in
X1. Therefore we disfavor AGN activity as the explanation for the
X-ray flare. Instead, it is a good soft TDE candidate. We only have
one X-ray detection (X1), which does not allow us to formally test
whether the event follows a $t^{-5/3}$ decline. However, we have a
deep observation C1 relatively close to X1 (i.e., 4.7 months before
X1), and the non-detection of the source in that observation allows us
to constrain the disruption time $t_\mathrm{D}$ to be close to or
after C1. In Figure~\ref{fig:lumlc}, we plot a possible
$(t-t_\mathrm{D})^{-5/3}$ decay curve, with $t_\mathrm{D}$ assumed to
be at the time of C1. The non-detection limits of our source in S1 and
X2, the latter of which is relatively deep, are consistent with this
curve.

We fortunately caught the X-ray flare in a deep observation (X1). We
detected short-term flux and spectral variability in X1, with the
X-ray spectrum being softer at lower flux. Based on our spectral
modeling, the variability could be due to change in the inner disk
truncation radius, by 24\% within $\sim$10 ks, at a constant mass
accretion rate. The variation timescale is thus 40 ks, close to the
viscosity timescale at the innermost stable circular orbit, which is
$\sim$70 ks, assuming a viscosity parameter of 0.1, a ratio of the
disk thickness to the inner disk radius of 0.1, and a BH mass of
$10^6$ \msun\ \citep[see Equation 5.69 in][]{frkira2002}. Therefore
the short-term variability in X1 could be due to viscosity instability
of the disk. Another soft TDE candidate \object{2XMMi
  J184725.1-631724} that we discovered in \citet{licagr2011} also
showed strong short-term variability in two deep \textit{XMM-Newton}
observations in the flare peak. We suggested the cause of its
short-term variability to be the fast change in the mass accretion
rate, considering the increase of the inner disk temperature with the
flux as inferred from the fits to a low-state spectrum and a
high-state spectrum in the second (brighter) \textit{XMM-Newton}
observation. We revisit this problem using the procedure as we adopted
for J1521+0749, by calculating the mass accretion rate directly
through ${\dot M}\propto T_\mathrm{MCD}^4N_\mathrm{MCD}^{3/2}$ for the
MCD when fitting the low-state and high-state spectra of the second
\textit{XMM-Newton} observation and fixing the absorption column
density at the value inferred from the fit to the average spectrum
from the whole observation (In \citet{licagr2011}, we allowed the
column density to be free but tied between the two spectra in fits;
fixing the column density instead allows to obtain better constraints
on the MCD parameters and thus more accurate comparison between the
two spectra). In this way the disk temperatures of the low-state and
high-state spectra were found to differ by 10 eV at the 7.9$\sigma$
confidence level, the inner disk radii by 17\% at the 3.3$\sigma$
confidence level, and the mass accretion rates by 1\%, less than
1$\sigma$ uncertainty. Therefore the short-term variability in
\object{2XMMi J184725.1-631724} could also be due to change in the
inner disk truncation radius at a constant mass accretion rate caused
by disk instability, as we suggest for J1521+0749. The TDE candidate
\object{SDSS J120136.02+300305.5} also showed short-term variability,
but its cause is not clear \citep{sarees2012}. Galactic stellar-mass
BH X-ray binaries in the thermal disk dominated state tend to show
very weak short-term variability \citep{remc2006}. The short-term
variability could be a unique phenomenon for soft TDEs.

We also detected some edge residuals in 0.5--1 keV when fitting the X1
spectrum with a MCD. One explanation for them is the presence of a
fast-moving warm absorber. This absorber could be the radiation-driven
outflow expected when the stellar debris fallbacks and accretes onto
the BH at a super-Eddington rate
\citep{re1988,stqu2009,stqu2011,loro2011}. In \object{SDSS
  J120136.02+300305.5}, \citet{sarees2012} also detected an edge, at
$\sim$0.66 keV, whose cause is not clear.

We did not detect UV or optical variability between X1 and
X2. Although we expect strong UV/optical emission from the disk and
the outflow in the flare peak, such emission could be still too faint
to be detected, maybe because our OM observations are not very deep
and/or the galactic stellar emission is too strong. Based on
\citet{loro2011}, the peak $g$-band (close to the OM $B$ filter)
emission has the AB magnitude about 22.3 mag for a TDE of
$M_\mathrm{BH}$ within $10^5$--$10^6$ \msun\ at the distance of
J1521+0749 (i.e., $D_L=866$ Mpc), while the galaxy has a $B$-band
magnitude of $21.2\pm0.4$ based on the X2 observation
(Table~\ref{tbl:xmmom}). The disk UV/optical emission can also be
estimated from modeling the X-ray spectrum of X1. For this, it is very
important to take into account the irradiation of the outer disk by
the inner disk emission. Therefore we adopted the irradiated disk
model \textit{diskir} in XSPEC \citep{gidopa2009}. We assumed pure
thermal disk emission, as the Compton tail was found to be very
weak. The reddening relation $\mathrm{E(B-V)}=1.7\times 10^{-22}
N_\mathrm{H}$ was adopted. The warm absorber was also required in the
fit, but it was assumed to be dust-free. The outer disk radius was set
to be 1000 times of the inner disk radius, which corresponds to
$\sim$100 times of the tidal disruption radius for a BH of $10^6$
\msun\ \citep{gumara2014}. The fraction of bolometric flux thermalized
in the outer disk $f_\mathrm{out}$ is assumed to be 0.1, which is
unfeasibly large \citep[$f_\mathrm{out}$ is typically $\sim$$10^{-3}$
  in stellar-mass BH X-ray binaries,][]{gidopa2009}. Even adopting the
above assumptions that favor the UV/optical emission, we still found
that such emission is about one order of magnitude fainter than the
galaxy emission (or the detection limit) in the OM filters used in X1
(Table~\ref{tbl:xmmom}).

Because of the super-Eddington accretion in TDEs, relativistic
outflows could be launched and interact with interstellar medium,
resulting in transient radio synchrotron emission that peaks at
$\sim$0.1--1 mJy (if the source is within Gpc distances) in $\sim$1 yr
after the disruption \citep{gime2011,bo2011}. However, despite
discovery of two jetted hard TDE candidates, there is no clear
evidence that soft TDEs also launch relativistic jets, because, thus
far, there is no radio detection that can be unambiguously ascribed to
a relativistic jet in soft TDEs \citep[see][for a review]{ko2015}. We
searched the Very Large Array (VLA) archival images and found only one
after 2000 April when our event occurred. It was taken on 2000 August
8 (close to X1 in time) in the L band (1.4 GHz). Our source was not
detected, with 3$\sigma$ upper limit about 0.8 mJy (three times the
rms value at the location of our source). The non-detection is not
constraining because the radio emission from the jet, even if it was
launched, should be still weak at the beginning of the flare
\citep{gime2011}.

The relatively large positional uncertainty of J1521+0749 does not
rule out that it resulted from disruption of a main-sequence star by a
massive BH residing in an off-nuclear (projected offset $<$2 kpc)
globular cluster in SDSS J152130.72+074916.5. However, this
explanation should not be very promising because if the BH mass is
$\sim$$10^5$--$10^6$ \msun\ as inferred from the X-ray spectral fit
(Section~\ref{sec:spmodel}) the cluster might have a much higher mass
\citep[e.g.,][]{pobahu2004} and thus would be very rare, especially
for a host that is not very massive. The rate of disrupting a normal
star by a massive BH in a globular cluster is also predicted to be
extremely low ($\lesssim$$10^{-7}$ yr$^{-1}$ per globular cluster),
especially at late times \citep[$>$2 Gyr,][]{bamaeb2004}.

The BH inferred for our source is relatively small but might still be
a little too large for the white dwarf (WD) tidal disruption, which
requires the BH mass to be $\lesssim$$2\times10^5$ \msun\ if the
disruption needs to be outside the event horizon or smaller if the
disruption is desired to be outside the innermost stable circular
orbit \citep[e.g.,][]{krpi2011}. The WD TDE is more prone to produce
an energetic jet, thus hard X-ray emission, than the main-sequence
star TDE, because the high density of WDs can support a strong
magnetic field on the BH \citep{krpi2011}. Even in the case of a slow
jet when the emission could be dominated by the jet photospheric
thermal emission, the spectrum is unlikely to be as soft as in
J1521+0749, due to Comptonization and broadening with pair production
\citep{shpere2013}. Based on the above considerations, we disfavor the
WD TDE explanation for J1521+0749.

The supersoft X1 spectrum of J1521+0749 is hardly seen in GRBs, even
for the small ultralong class. Only the ultralong GRB 060218 showed a
similarly soft spectrum \citep[photon index
  $\sim$5.5,][]{magula2015}. The main problem of associating
J1521+0749 with an ultralong GRB is that its host is consistent with
an early-type galaxy, while most ultralong GRBs have hosts showing
intensive star forming activity \citep[e.g.,][]{letast2014}. Our
source is unlikely to be a SN, because SNe hardly have
$L_\mathrm{X}>10^{42}$ erg~s$^{-1}$ and supersoft X-ray
spectra. While the shock breakout in SN 2008D had peak
  $L_\mathrm{X}$ similar to J1521+0749 and showed relatively soft spectra
  \citep[maximum photon index $\sim$3.2,][]{sobepa2008}, the event was
  short ($\lesssim600$ s) and had a fast rise, exponential decay
  profile. In contrast, the X1 observation of J1521+0749 was long
  ($\sim$13 hrs) and had much more steady count rates (the second half of the
  observation had 0.2--2 keV pn count rates higher by 60\% than those
  in the first half only). Although the shock breakout in
  a massive star exposion might last for hours \citep{nasa2012}, the
  X1 observation of J1521+0749 was still too long and had too steady
  count rates to be part of such an event; the old stellar
  environment of J1521+0749 also argues against the explanation of a
  massive star explosion.

J1521+0749 appears in the direction of the cluster of galaxies
\object{MKW 3s} ($z=0.045$). Its candidate host galaxy \object{SDSS
  J152130.72+074916.5} is not a member of this cluster. J1521+0749 is
relatively close to, but still too far away to be in, two galaxies in
the cluster, \object{SDSS J152131.80+074851.9} ($z=0.04533$ or
$d_L=201$ Mpc) and \object{SDSS J152134.15+074815.3} ($z=0.04225$ or
$d_L=187$ Mpc, see Figure~\ref{fig:cfhtimg}).
We can rule out J1521+0749 as a coronally active star based on the
ratio of the 0.2-12 keV maximum flux to the $K$-band flux, which is
$\log(F_\mathrm{X}/F_\mathrm{IR})=1.05$ for J1521+0749, much higher
than seen in coronally active stars
\citep[$\log(F_\mathrm{X}/F_\mathrm{IR})\lesssim-0.9$,][]{liweba2012}. The
ultrasoft spectrum makes J1521+0749 similar to supersoft X-ray sources
\citep[SSS;][]{kava2006,gr2000}. They are mostly cooling white dwarfs
\citep[WDs;][]{kava2006} or nuclear burning of the hydrogen-rich
matter on the surface of a WD in the so-called close binary supersoft
sources or supersoft novae \citep{kava2006,gr2000}. J1521+0749 is
unlikely to be a cooling WD due to its large variability
\citep{yulitr1996}. It is unlikely to be a close binary
supersoft source or a supersoft novae either, because such objects are
uncommon \citep[only a few tens found in our Galaxy,
][]{kava2006,gr2000} and the chance to find one consistent with the
center of a bright galaxy is essentially zero.

Therefore J1521+0749 is best explained as a soft TDE. Its high-quality
X-ray data in X1 have allowed us to search for short-term variability
and carry out detailed spectral modeling, which revealed rich
information of the accretion process in the event. Few TDE candidates
have such high-quality X-ray data. More soft TDEs of similar quality
are needed to check whether the short-term variability and the edge
that we detected in J1521+0749 are common in such events and to
understand their origins.

\acknowledgments 
We thank the anonymous referee for the helpful
comments. We thank Andrej Lobanov for a critical reading of the
manuscript. This work is partially supported by NASA ADAP Grant
NNX10AE15G.

Funding for SDSS-III has been provided by the Alfred P. Sloan Foundation, the Participating Institutions, the National Science Foundation, and the U.S. Department of Energy Office of Science. The SDSS-III web site is http://www.sdss3.org/. SDSS-III is managed by the Astrophysical Research Consortium for the Participating Institutions of the SDSS-III Collaboration including the University of Arizona, the Brazilian Participation Group, Brookhaven National Laboratory, Carnegie Mellon University, University of Florida, the French Participation Group, the German Participation Group, Harvard University, the Instituto de Astrofisica de Canarias, the Michigan State/Notre Dame/JINA Participation Group, Johns Hopkins University, Lawrence Berkeley National Laboratory, Max Planck Institute for Astrophysics, Max Planck Institute for Extraterrestrial Physics, New Mexico State University, New York University, Ohio State University, Pennsylvania State University, University of Portsmouth, Princeton University, the Spanish Participation Group, University of Tokyo, University of Utah, Vanderbilt University, University of Virginia, University of Washington, and Yale University.

\end{document}